\documentclass[]{interact}

\usepackage{epstopdf}

\usepackage{caption}
\usepackage{subcaption}

\usepackage[numbers,sort&compress]{natbib}
\bibpunct[, ]{[}{]}{,}{n}{,}{,}
\renewcommand\bibfont{\fontsize{10}{12}\selectfont}

\theoremstyle{plain}

\theoremstyle{definition}

\theoremstyle{remark}

\usepackage{lineno,hyperref}

\usepackage{color,soul}

\makeatletter
\AtBeginDocument{\let\hl\@firstofone}
\makeatother

\begin{document}

\articletype{Full length article : Phil.Mag.}

\title{Role of axial twin boundaries on deformation mechanisms in Cu nanopillars}

\author{
\name{P. Rohith\textsuperscript{a,b}, G. Sainath\textsuperscript{b}\thanks{CONTACT G. Sainath. Email: sg@igcar.gov.in/mohansainathan@gmail.com}, 
Sunil Goyal\textsuperscript{a,b}, A. Nagesha\textsuperscript{a,b}, and V.S. Srinivasan\textsuperscript{a,c}}\affil{\textsuperscript{a}Homi 
Bhabha National Institute (HBNI), 
Indira Gandhi Centre for Atomic Research, Kalpakkam, Tamilnadu-603102, India;\\ 
\textsuperscript{b}Materials Development and Technology Division, Indira Gandhi Centre for Atomic Research, Kalpakkam, Tamilnadu-603102, 
India;\\
\textsuperscript{c}Scientific Information Resource Division, Resource Management \& Public Awareness Group, Indira Gandhi Centre for Atomic 
Research, Kalpakkam, Tamilnadu-603102, India}
}

\maketitle

\begin{abstract}
In recent years, twinned nanopillars have attracted tremendous attention for research due to their superior mechanical properties. However, 
most of the studies were focused on nanopillars with twin boundaries (TBs) perpendicular to loading direction. Nanopillars with TBs parallel 
to loading direction have received minimal interest. In this backdrop, the present study is aimed at understanding the role of axial TBs 
on strength and deformation behaviour of Cu nanopillars using atomistic simulations. Tensile and compression tests have been performed on 
$<$112$>$ nanopillars with and without TBs. Twinned nanopillars with twin boundary spacing in the range 1.6-5 nm were considered. The 
results indicate that, under both tension and compression, yield strength increases with decreasing twin boundary spacing and is always 
higher than that of perfect nanopillars. Under compression, the deformation in $<$112$>$ perfect as well as twinned nanopillars proceeds 
by the slip of extended dislocations. In twinned nanopillars, an extensive cross-slip by way of Friedel-Escaig and Fleischer mechanisms has 
been observed in compression. On the other hand, under tensile loading, the deformation in perfect nanopillars occurs by partial slip/twinning, 
while in twinned nanopillars, it proceeds by the slip of extended dislocations. This extended dislocation activity is facilitated by 
stair-rod formation and its dissociation on the twin boundary. Similar to compressive loading, the extended dislocations under tensile 
loading also exhibit cross-slip activity in twinned nanopillars. However, this cross-slip activity occurs only through Fleischer mechanism 
and no Friedel-Escaig mechanism of cross-slip has been observed under tensile loading.
\end{abstract}

\begin{keywords}
Atomistic simulations; Nanopillars; Twin boundaries; Dislocations; Cross-slip
\end{keywords}



 

\section{Introduction}
In the last couple of years, the twinned nanowires/nanopillars received tremendous attention for research due to their superior 
mechanical performances. Twinned nanowires exhibit enhanced strength without loss of ductility \cite{Cao2007, Lu2009}, improves 
crack resistance \cite{Liu2014}, and show high strain rate sensitivity \cite{Deng2010}. The twin boundaries in nanowires can also 
acts as a source/sink and also as a glide plane for dislocations \cite{Sun2011}. In nanowires, the presence of twin boundaries 
influences the plastic flow by way of strain hardening or strain softening or in some cases, without any hardening or softening, 
which depends on twin boundary orientation, twin boundary spacing, nanowire size and shape \cite{Deng2009,Zhang2009,Zhu2011,Sun2017}. 
For example, Zhang and Huang \cite{Zhang2009} reported that in nanowires, the strengthening or softening due to twin boundaries 
depends on the cross-section shape of the nanowire. In a square cross-section nanowire, the twin boundaries strengthen the nanowire, 
while in circular cross-section nanowire, the presence of twin boundaries leads to softening \cite{Zhang2009}. \hl{This difference 
under the influence of twin boundaries is mainly due to differences in stress required for dislocation nucleation during yielding 
in square and circular nanowires \mbox{\cite{Zhang2009}}. For nanowires with square cross-section, the presence of sharp corners 
lowers the stress required for dislocation nucleation, which is lower than that needed for dislocation penetration through the 
twin boundaries. This leads to strengthening in square cross-section nanowires. On the other hand, the nucleation stress in 
circular nanowires is higher, so the penetration through the twin boundaries requires no additional stress leading to the observed 
softening in circular nanowires}. Similarly, Sun et al. \cite{Sun2017} have investigated the combined influence of twin boundary 
spacing and nanowire length. It has been reported that long nanowires with length (L) $>$ 450 nm always exhibit brittle failure, 
while short nanowires with L $<$ 200 nm invariably fails in ductile manner, irrespective of twin boundary spacing. However, the 
nanowires with intermediate lengths (200 nm $<$ L $<$ 450 nm) have shown a ductile to brittle transition with increasing twin 
boundary spacing (high ductility for low twin boundary spacing) \cite{Sun2017}. 

Apart from influencing the strength and ductility, the twinned nanopillars have shown novel deformation mechanisms 
\cite{Sun2011,Wang2006,Wang2009,Jang2012}. Generally, dislocations in FCC materials such as Cu glide on $\{111\}$ planes. However, 
in twinned nanowires, the $1/2<110>$ dislocations glide on $\{100\}$ plane after penetrating the twin boundary \cite{Wang2006}. 
In another study, Wang and Sui \cite{Wang2009} have shown that the leading and trailing partials can exchange their order after 
passing through the twin boundary. Using atomistic simulations and in-situ experiments, Jang et al. \cite{Jang2012} reported a 
brittle to ductile transition in orthogonally twinned Cu nanopillars with decreasing twin boundary spacing. Further, the twin 
boundary orientation also affects the deformation mechanisms in FCC nanopillars. Due to special geometry, the twin boundaries 
orientated at different angles with respect to the loading direction possess distinct interactions with dislocations, which 
results in different deformation mechanisms \cite{Jang2012,SainathPLA2015,Yang2017,Sun2018}. In nanopillars with orthogonal 
twin boundaries (twin boundaries perpendicular to the loading direction), the twin boundary-dislocation interactions dominate 
the plastic flow and increases the flow stress, whereas de-twinning governs deformation in nanopillars with slanted twin boundaries 
(twin boundaries oriented other than $90^o$ and $0^o$ with respect to loading direction) \cite{Jang2012}. \hl{In nanopillars with 
orthogonal twin boundaries, different dislocation-twin interactions such as dislocation transmission across the twin boundary, 
stair-rod formation and dislocation multiplication have been observed. Afanasyev and Sansoz \mbox{\cite{Afanasyev2007}} have 
reported the formation and dissociation of Lomer-Cottrell locks during the dislocation-twin boundary interactions in Au nanopillars. 
Further, the dislocations with Burgers vector parallel to the twin plane, can either propagate into the adjacent twin grain by 
cutting through the boundary, or be absorbed and dissociate within the boundary plane \mbox{\cite{Afanasyev2007,Kulkarni}}. 
Similarly, Jang et al. \mbox{\cite{Jang2012}} have reported that the dislocation-twin boundary interactions leads to entanglement 
and multiplication without any change in twin boundary spacing in orthogonally twinned nanopillars. In contrary to orthogonally 
twinned nanopillars, the motion of Shockley partials along the twin boundary leading to twin boundary migration and de-twinning 
has been observed in nanopillars with slanted twin boundaries \mbox{\cite{Jang2012,Sun2018}}}.

Most of the studies mentioned above were focused on the nanowires with orthogonal or slanted twin boundaries. Little has been 
investigated on twinned nanowires when the twin boundaries are parallel to the loading direction (axial or longitudinal twin 
boundaries). \hl{In longitudinally twinned nanowires, mainly two loading directions ($<$110$>$ and $<$112$>$) have been investigated 
in the literature, with more focus on $<$110$>$}. In $<$110$>$ oriented longitudinally twinned FCC nanowires, it has been shown 
that the introduction of twin boundary changed the deformation mode from twinning in perfect nanowires to slip in twinned nanowires 
\cite{SainathPLA2015,Roos-APL}. Similarly, Jeon and Dehm \cite{Jeon-Dehm} have observed the formation of intensive dislocation 
networks on the axial twin boundary under both $<$110$>$ or $<$112$>$ loading directions. Further, the type of dislocation network 
strongly depends on the nanowire orientation. In $<$110$>$ loading direction, a high density of sessile dislocation networks formed 
at the twin boundary, while in $<$112$>$ loading direction, a lower density of glissile dislocation networks was observed 
\cite{Jeon-Dehm}. In nanopillar containing a single axial twin boundary, Cheng et al. \cite{Cheng-PRL} reported the complete 
annihilation of twin boundary leading to the formation of a single crystal, which has been attributed to anomalous de-twinning 
mechanism. However, the influence of twin boundary spacing has not been investigated when the twin boundaries are parallel to 
the loading direction. Although, under tensile loading, it has been shown that the introduction of longitudinal twin boundary 
increases the strength of the nanowire compared to its perfect counterpart \cite{SainathPLA2015}, but the influence of twin boundary 
spacing is still elusive. Further in twinned nanopillars, the dislocation interactions with twin boundaries needs to be investigated 
in greater detail. For example, how the dislocations transmit/cross-slip across/onto twin boundaries is still ambiguous \cite{Zhu2011}. 
With these multiple motivations, an attempt has been made in this paper to understand the spacing dependence of yield strength, 
deformation mechanisms, different dislocation-twin interactions and possible dislocation cross-slip mechanisms under tensile and 
compressive loading of $<$112$>$ oriented longitudinally twinned Cu nanopillars. The twin boundaries have been introduced parallel 
to the loading direction ($<$112$>$) and number of twin boundaries varied from 1 to 5, which results in twin boundary spacing in 
the range 1.6-5 nm. 

\section{Computational details}

Molecular dynamics (MD) simulations have been performed using Large scale Atomic/Molecular Massively Parallel Simulator (LAMMPS) 
package \cite{Plimpton-1995} employing an embedded atom method (EAM) potential for FCC Cu developed by Mishin and co-workers 
\cite{Mishin-2001}. This potential has been chosen for being able to reproduce generalized stacking fault and twinning fault energies 
for Cu \cite{Liang-PRB}, which are key variables for predicting the dislocation and plasticity related properties. AtomEye \cite{AtomEye} 
and OVITO \cite{Ovito} packages have been used for the visualisation and analysis of deformation mechanisms. Common neighbour analysis 
(CNA) as implemented in AtomEye and OVITO and dislocation extraction algorithm \cite{DXA} as implemented in OVITO have been used to 
identify the stacking faults and various type of dislocations based on the Burgers vector.   

\begin{figure}[h]
\centering
\includegraphics[width=7.5cm]{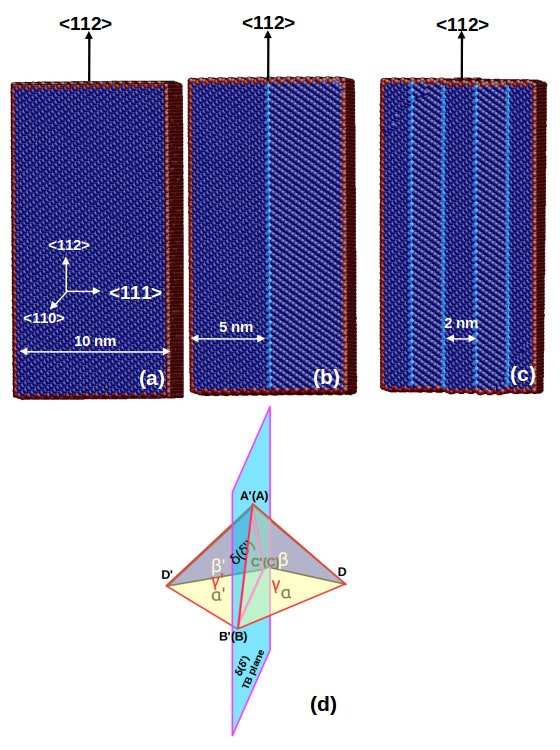}
\caption {\hl{The model system of (a) perfect and, twinned Cu nanopillars with (b) one and (c) four equidistant twins. The double 
Thompson tetrahedron illustrating the slip planes and directions is shown in (d). The crystallographic orientation and twin boundary 
spacing has been shown for clarity. The cross-section width of all the nanowires is 10 nm. In (a)-(c), the front surfaces are removed 
and the atoms are coloured according to the common neighbour analysis (CNA)}.}
\label{Initial}
\end{figure}

In order to create a longitudinal twin boundary, the procedure followed in Ref. \cite{Sainath-PhilMag} has been adapted. First, a 
defect free Cu nanopillar of square cross section width (d) = 10 nm, oriented in $<$112$>$ axial direction with \{110\} and \{111\} 
as side surfaces has been chosen. The nanopillar length (l) was twice the cross section width (d) and no periodic boundary conditions 
were used in any direction. Then, the nanowire has been divided into two equal parts along the $<$111$>$ direction and, one part of 
the crystal is rotated with respect to the other by $180^o$. Following the rotation, a twin boundary forms at the interface, which 
lies on \{111\} plane. Similar procedure has been adapted to create more number of twin boundaries. The nanopillars containing one, 
two, three, four and five twin boundaries resulted in twin boundary spacings of 5.0, 3.3, 2.5, 2.0 and 1.6 nm, respectively. The 
typical nanopillars with different twin boundary spacings considered in this study along with perfect nanopillar are shown in 
Figure \ref{Initial}. \hl{The double Thompson tetrahedron showing the slip planes and directions in parent and twinned lattice is 
shown in Figure \mbox{\ref{Initial}d}. In the present study, this notation has been used to describe the various dislocation-twin 
boundary interactions.}

Following the creation of the model nanowires, the energy minimization was performed by a conjugate gradient method to obtain a 
relaxed structure. To put the sample at the required temperature, all the atoms have been assigned initial velocities according to
the Gaussian distribution. Following this, the nanopillar system was thermally equilibrated to a required temperature of 10 K in 
canonical ensemble (constant NVT) with a Nose-Hoover thermostat. The velocity verlet algorithm has been used to integrate the 
equations of motion with a time step of 2 fs. Following thermal equilibration, the tensile or compressive deformation was carried out 
in a displacement controlled mode at a constant strain rate of $1 \times 10^{8}$ s$^{-1}$ by imposing displacements to atoms along the axial 
direction (i.e., $<$112$>$ axis) that varied linearly from zero at the bottom to a maximum value at the top layer. The strain 
($\varepsilon$) has been calculated as $(l-l_0)/l_0$, where $l$ is instantaneous length and $l_0$ is the initial length of the 
nanowire. Finally, the stress has been obtained using the Virial expression \cite{Virial}, which is equivalent to a Cauchy's stress 
in an average sense.

\section{Results}

\subsection{Effect of twin boundaries on stress-strain behaviour and yield stress}

\begin{figure}
\centering
\includegraphics[width=14cm]{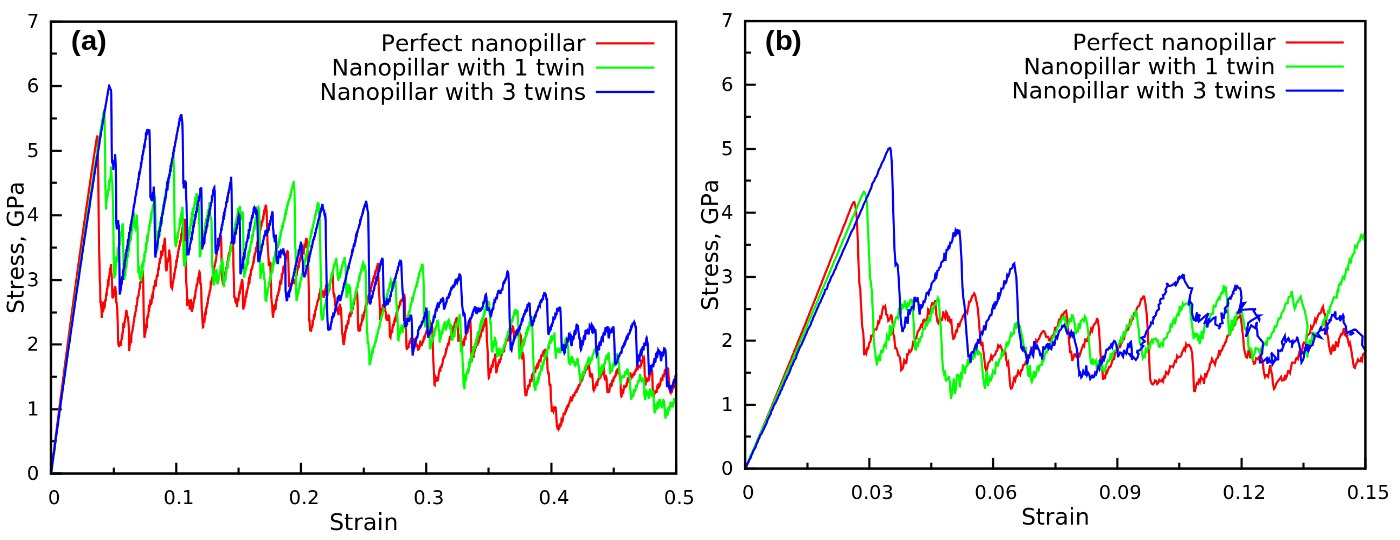}
\caption {The stress-strain curves of $<$112$>$ Cu nanopillars with and without axial twin boundaries under (a) tension (b) compression.}
\label{Stress-strain}
\end{figure}

Figure \ref{Stress-strain} shows the stress-strain behaviour of perfect as well as twinned $<$112$>$ Cu nanopillars subjected to tensile 
and compressive loading. The tensile loading has been performed until a nanopillar attains a strain of 50\%, while in compression, 
the loading has been applied until a strain of 15\%. It can be seen that, all the nanopillars exhibits a linear elastic deformation 
up to a maximum value of stress (designated as yield stress) followed by a precipitous drop of around 2-3 GPa. Following the drop, 
the flow stress under tensile loading show an overall decreasing behaviour with large fluctuations (Figure \ref{Stress-strain}a). On 
the other hand, the flow stress under compressive loading display large fluctuations around a constant mean stress of 2 GPa (Figure 
\ref{Stress-strain}b). The peak value of the stress-strain curve has been taken as the yield stress of the nanopillar. It can be seen 
that, under both tensile and compressive loading, the introduction of twin boundaries increases the strength of the nanopillar as 
compared to the perfect nanopillar (Figure \ref{Stress-strain}). Further, higher the number of twin boundaries, higher the strength. 

\begin{figure}[h]
\centering
\includegraphics[width=14cm]{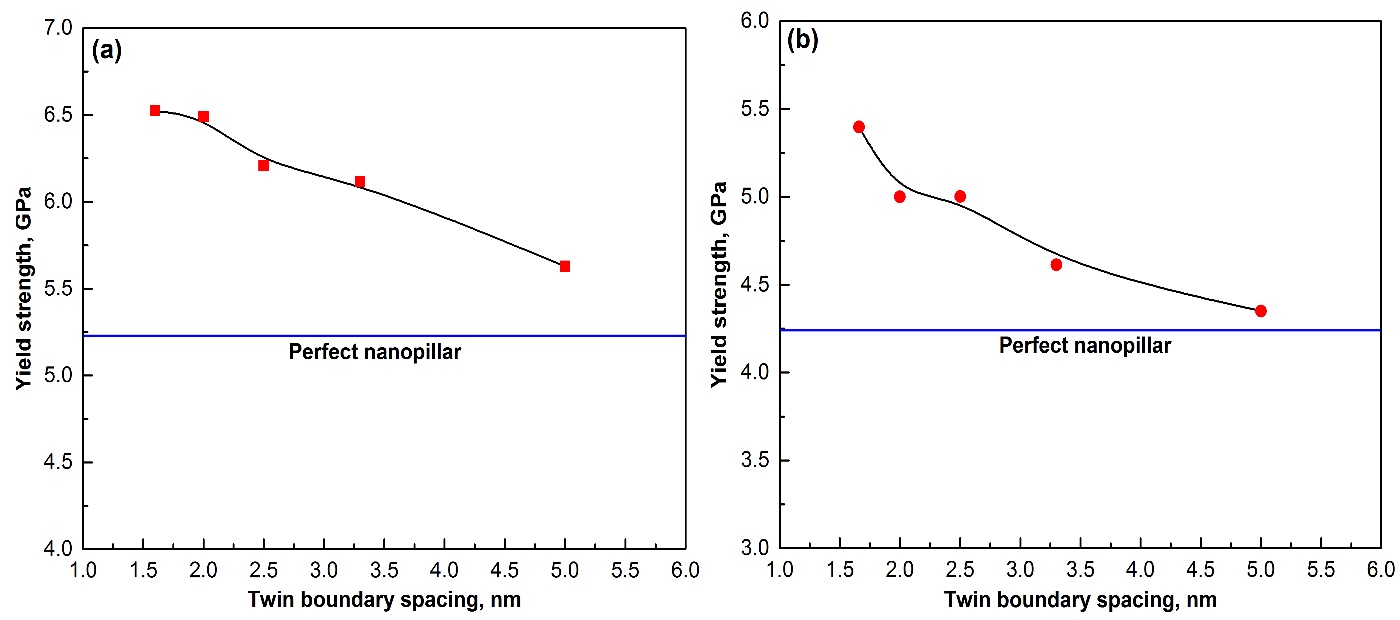}
\caption {The variation of yield strength as a function of twin boundary spacing (TBS) under (a) tensile (b) compressive loading. 
For reference, the yield strength of perfect nanopillar has been represented as a horizontal line.}
\label{Fig03}
\end{figure}

\begin{figure}
\centering
\includegraphics[width=7.5cm]{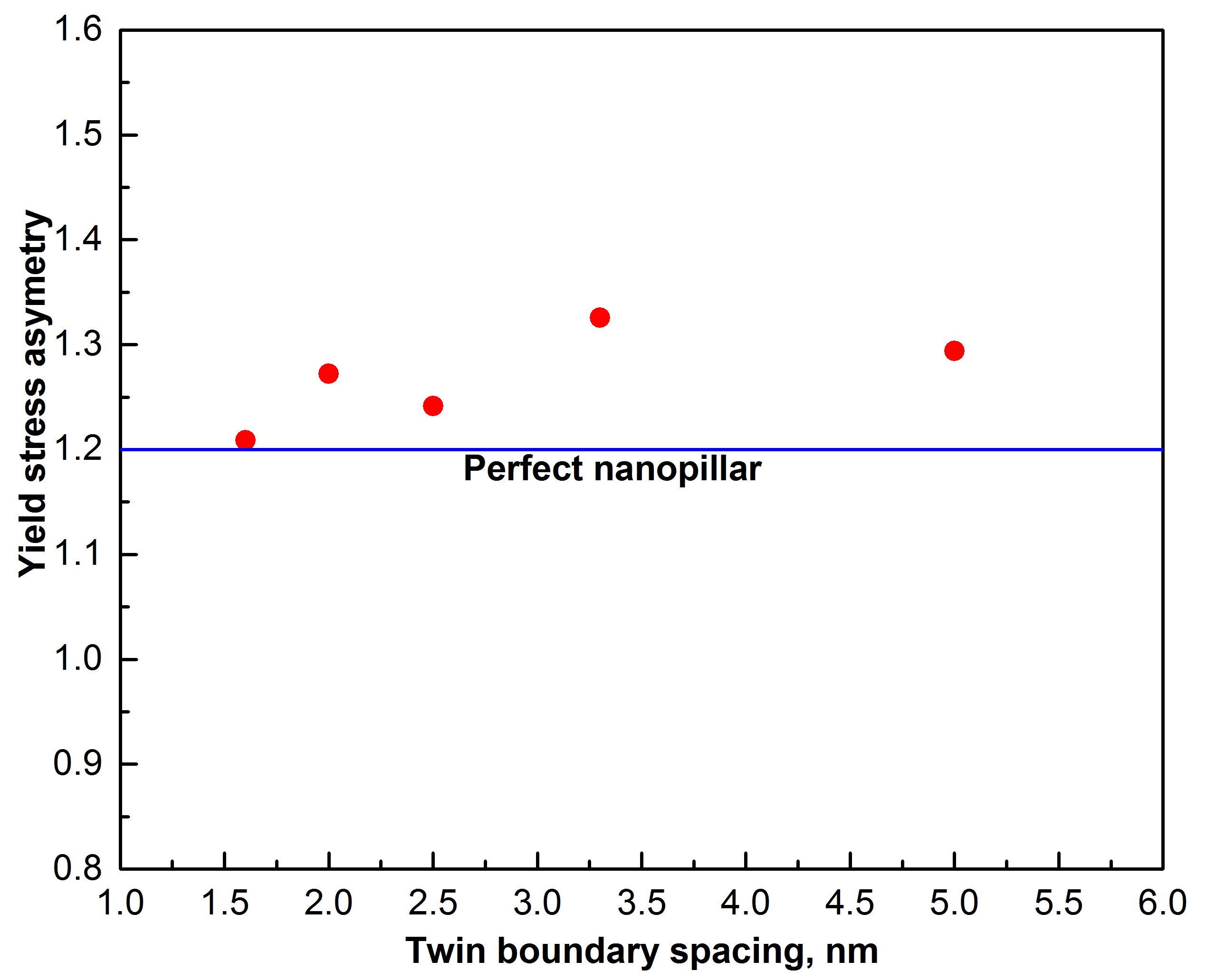}
\caption {Variation of yield strength asymmetry in twinned nanopillar of different twin boundary spacings. For comparison, the yield 
strength asymmetry of perfect nanopillar has been shown as a horizontal line. \hl{Here, the yield strength asymmetry is defined as the 
ratio of yield stress in tension ($\sigma_{yt}$) to that in compression ($\sigma_{yc}$) }.}
\label{Fig04}
\end{figure}

To quantify the effect of twin boundaries, the twin boundary spacing has been chosen as the 
variable parameter. Figure \ref{Fig03} shows the variation of yield stress as a function of twin boundary spacing under tensile 
and compressive loading. For comparison, the yield stress of perfect $<$112$>$ nanopillar has been shown as a horizontal line 
under respective loading conditions. It can be seen that the yield stress of all the twinned nanopillars falls above the horizontal 
line (Figure \ref{Fig03}), indicating that the strength of twinned nanopillars is higher the perfect twin free nanopillars. Further, 
the yield stress values under tensile and compressive loading decreases with increasing twin boundary spacing (Figure \ref{Fig03}), 
which is similar to that observed in orthogonally twinned FCC nanopillars \cite{Cao2007,Zhang2009,Jang2012, Yang2017,Afanasyev2007}. 
It can also be seen 
that, for perfect as well as twinned nanopillars, the yield stress values under tensile loading are higher than that in compressive 
loading i.e., all the nanopillars display tension-compression asymmetry in yield stress. Figure \ref{Fig04} shows the values of 
yield stress asymmetry measured as the ratio of yield stress in tension to that in compression, as a function of twin boundary 
spacing. The yield stress asymmetry of perfect nanopillar is obtained as 1.2, which is shown as horizontal line in Figure \ref{Fig04}. 
It can be seen that the yield stress asymmetry for twinned nanopillars is different from that of the perfect nanopillars (higher 
the number of twin boundaries, lower the asymmetry), which indicates that the twin boundaries can also influence the tension-compression 
asymmetry. Previous studies have shown that the asymmetry in yield stress arises due to different values of Schmid factor for leading 
partials under tensile and compressive loading \cite{Rohith-ComCondMater}. Further, Salehinia and Bahr \cite{Salehinia-IJP} have shown that 
the asymmetry in the strength of nanowires decreases with increasing defect density, which is in agreement with the present findings.

\subsection{Deformation mechanisms in perfect $<$112$>$ Cu nanopillars}

\begin{figure}
\centering
\includegraphics[width=10cm]{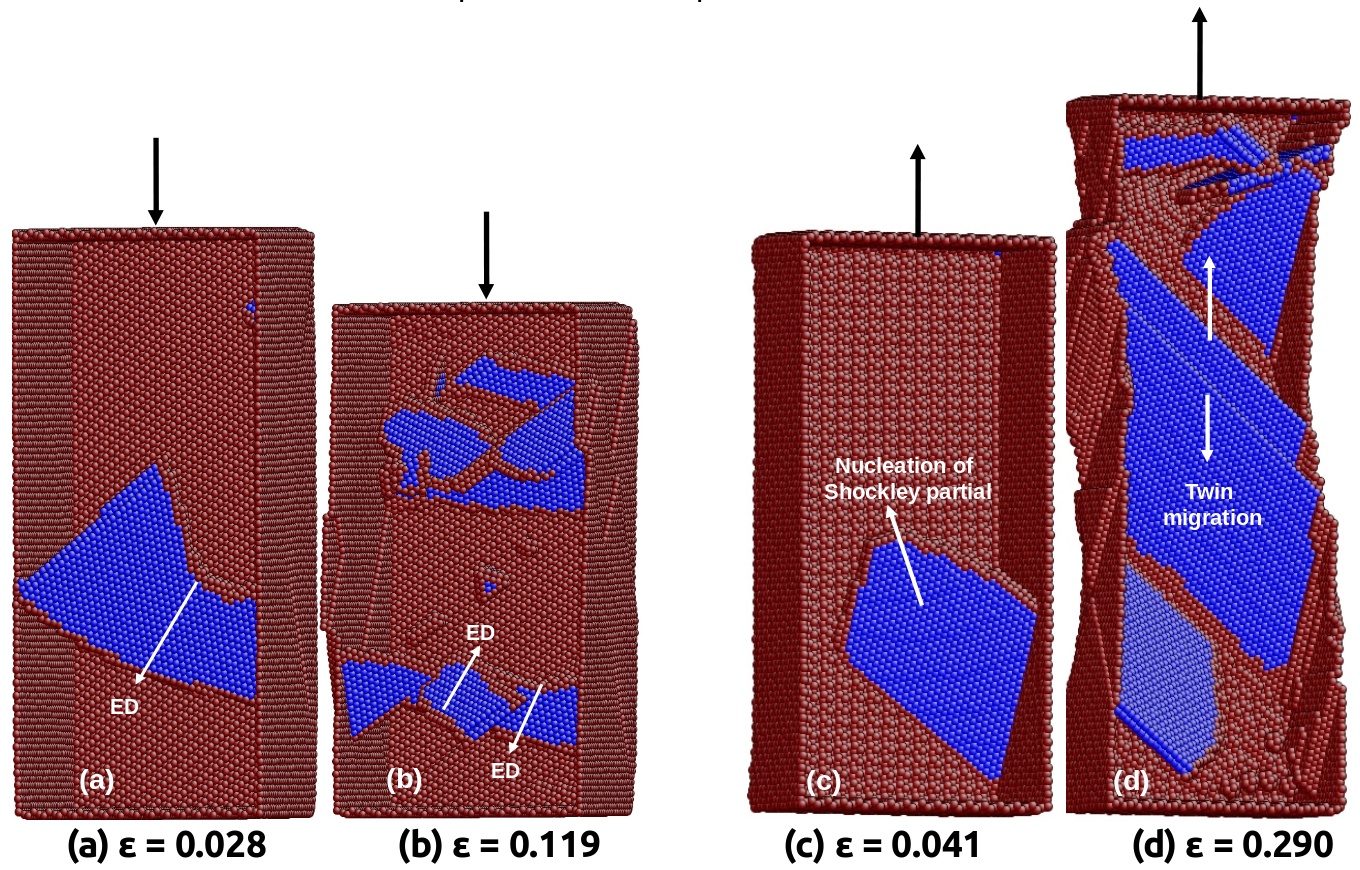}
\caption {The atomic snapshots displaying the nucleation and the dominant deformation behaviour under (a) \& (b) compressive, 
and (c) \& (d) tensile loading. The deformation is dominated by extended dislocation \hl{(ED)} under compressive loading, and slip through 
partial dislocations/twinning under tensile loading. The atoms are coloured according to the common neighbour analysis (CNA). The 
blue colour atoms represent the FCC and the red colour atoms indicate surfaces and dislocation core. The white arrows indicate the 
direction of dislocation/twin boundary motion.}
\label{Fig05}
\end{figure}

In order to understand the clear role of twin boundaries on deformation mechanisms, first, the deformation mechanisms in perfect 
$<$112$>$ Cu nanopillars has been investigated. Figure \ref{Fig05} shows the deformation behaviour in perfect $<$112$>$ Cu nanopillars 
under compressive and tensile loading. As shown in Figure \ref{Fig05}(a-b), the deformation is governed by the slip of extended 
dislocations under compressive loading. An extended dislocation consist of a leading and trailing partial dislocations separated by 
a stacking fault. Figure \ref{Fig05}a shows an extended dislocation nucleated during the yielding of Cu nanopillars under compressive 
loading. With increasing strain, the extended dislocations belonging to different slip systems interact and results in the formation 
of a stacking fault tetrahedron (Figure \ref{Fig05}b). The nucleation of trailing partials under compressive loading is favoured due 
to its high Schmid factor value as compared to leading partial \cite{Rohith-ComCondMater}. Different from compressive loading, the 
deformation under tensile loading of perfect $<$112$>$ Cu nanopillars occurs through the slip of Shockley partial dislocations 
(Figure \ref{Fig05}(c-d)). The yielding occurs through the nucleation of a $1/6<112>$ Shockley partial dislocations (Figure 
\ref{Fig05}c). Since the nanopillar is defect-free, the nucleated partial dislocation glides freely towards the opposite surface 
leaving behind a stacking fault, and also a step on the surface of nanopillar. The continuous nucleation and glide of Shockley 
partials on the adjacent parallel planes leads to the formation of a micro-twin. With increasing strain, this micro-twin grows into 
complete twin as shown in Figure \ref{Fig05}d. Overall, the deformation under tensile loading of perfect $<$112$>$ Cu nanopillars 
is dominated by the slip of partial dislocations along with deformation twinning, which is in agreement with that predicted using 
Schmid factor analysis \cite{Rohith-ComCondMater}. Interestingly, the nucleation of trailing partials has not been observed under 
tensile loading.

\subsection{Deformation mechanisms in twinned $<$112$>$ Cu nanopillars under compressive loading}

\begin{figure}
\centering
\includegraphics[width=12cm]{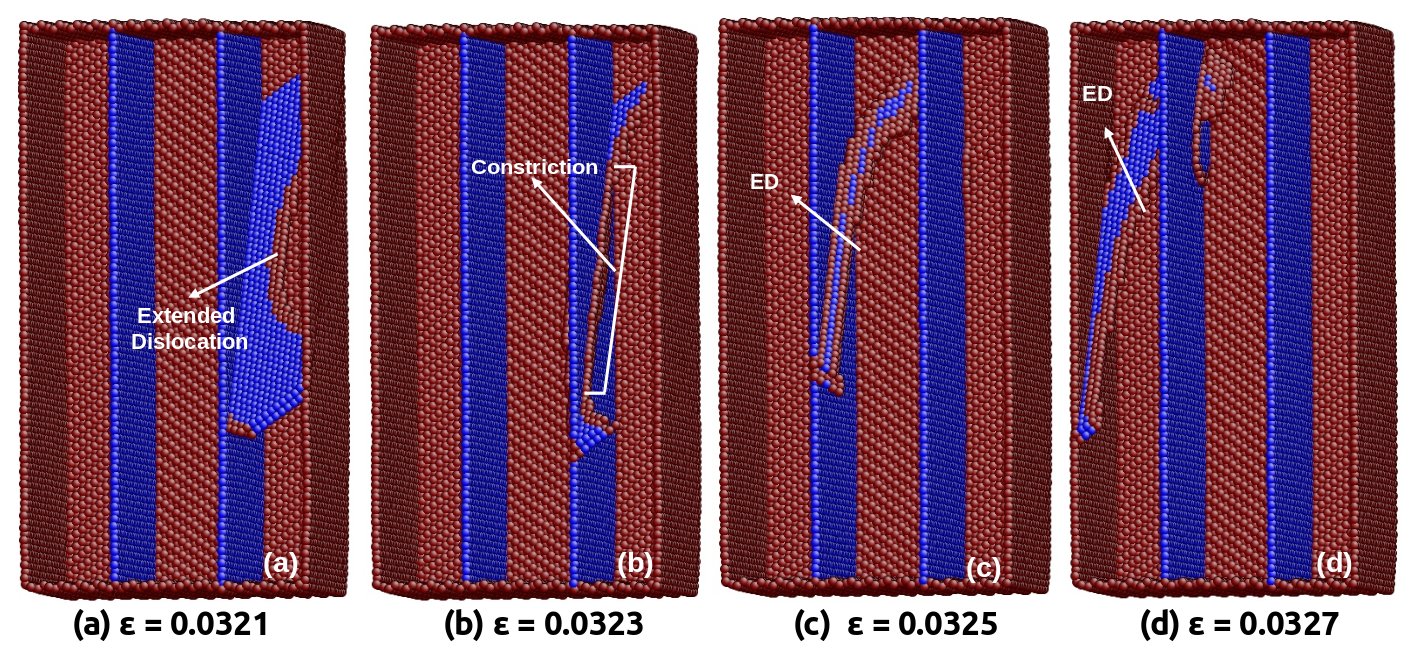}
\caption {The deformation behaviour under the compressive loading of twinned Cu nanopillars at different strains. The glide of an extended 
dislocation \hl{(ED)} undergoing slip transmission through multiple twin boundaries can be seen. This process involves the constriction of an extended 
dislocation followed by cross-slip. The atoms are coloured according to the common neighbour analysis (CNA). The blue colour atoms 
represents the FCC and the red colour atoms indicate the surfaces and dislocation core.}
\label{112-compression}
\end{figure}

The deformation of Cu nanopillar with single and multiple twin boundaries has been investigated under compressive loading. As a 
representative one, the deformation behaviour of Cu nanopillar with two twin boundaries has been shown in Figure \ref{112-compression}. 
It can be seen that the yielding in twinned nanopillars occurs through the nucleation of extended dislocation consisting of leading 
(1/6[$\bar1$12]) and trailing (1/6[121]) partials. Following yielding, the extended dislocation glides towards a nearby twin boundary 
as shown in Figure \ref{112-compression}a. Since the presence of twin boundary restricts the free glide, the extended dislocation gets 
constricted and forms full dislocation at the twin boundary (Figure \ref{112-compression}b) according to the following reaction: 
\begin{equation} 
\label{eq:1}
{1\over6} [\bar{1}12] + {1\over6}[121] \longrightarrow {1\over2} [011] \quad \text{or} \quad B\alpha + \alpha C \longrightarrow BC
\end{equation}This constricted full dislocation, which has high energy, cross-slips to neighbouring twinned grain on a plane symmetric 
to the original (symmetric slip transmission) by dissociating into an extended dislocation (Figure \ref{112-compression}c). This 
dissociation can be written as 
\begin{equation} 
\label{eq:2}
{1\over2} [011]  \longrightarrow {1\over6} [\bar{1}12]_{T} + {1\over6}[121]_{T} + \text{SF} \quad \text{or} \quad BC(B'C') \longrightarrow B'\alpha ' + \alpha 'C'
\end{equation}Here, the subscript "T" for the leading and trailing partial indicates Burger's vector in twinned grain. The extended 
dislocation again undergoes similar constriction and transmission upon facing the subsequent twin boundaries before annihilating at 
the free surface (Figure \ref{112-compression}d).

\begin{figure}[h]
\centering
\includegraphics[width=10cm]{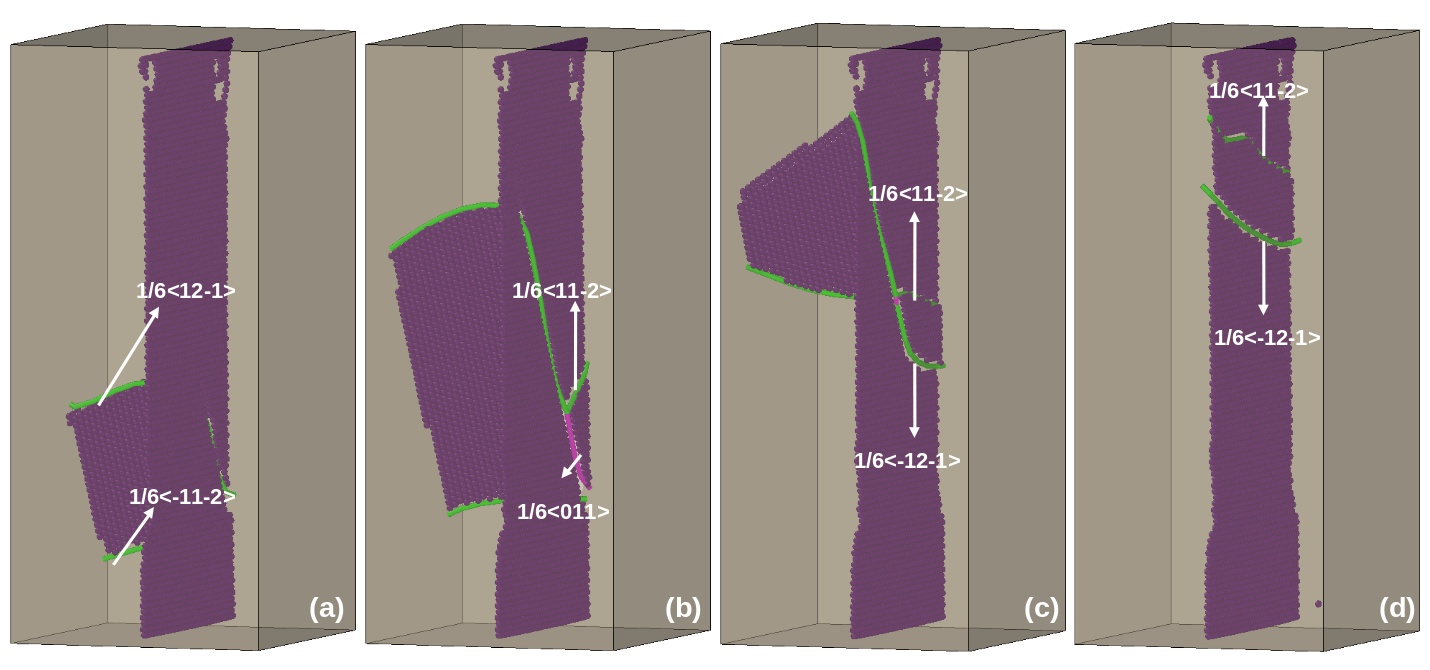}
\caption {The deformation behaviour under the compressive loading of twinned Cu nanopillars (a-d). Here, the extended dislocation 
cross-slips to the twin boundary plane by forming an intermediate stair-rod dislocation. The purple colour surface represents 
the stacking fault and twin boundary. The green colour lines indicate the Shockley partial dislocations and the magenta colour 
line represents stair-rod dislocation.}
\label{112-compression1}
\end{figure}

In addition to symmetric slip transmission aided by the formation of constriction and dissociation, it has been observed that the 
extended dislocation can also cross-slip to twin boundary plane without forming any intermediate constriction (Figure \ref{112-compression1}). 
Initially, the glide of leading partial (part of an extended dislocation) is restricted at the twin boundary with trailing partial far 
behind (Figure \ref{112-compression1}a). This leading partial with Burger vector of 1/6[12$\bar1$] dissociates into a stair-rod 
dislocation (1/6[011]) and a Shockley partial dislocation 1/6[11$\bar{2}$] on the plane of the twin boundary (Figure \ref{112-compression1}b). 
In Burger vector terms, this dissociation can be written as
\begin{equation} 
\label{eq:3}
{1\over6} [12\bar{1}] \longrightarrow {1\over6}[11\bar{2}]+{1\over6}[011] \quad \text{or} \quad C\beta \longrightarrow C\delta + \delta \beta
\end{equation}With increasing deformation, the trailing partial comes to the twin boundary and its interaction with existing 
stair-rod dislocation leads to the formation of Shockley partial lying on the twin boundary (Figure \ref{112-compression1}c). This 
reaction can be written as
\begin{equation} 
\label{eq:4}
{1\over6}[011] + {1\over6}[\bar{1}1\bar{2}] \longrightarrow {1\over6}[\bar{1}2\bar{1}] \quad \text{or} \quad \delta \beta + \beta A \longrightarrow \delta A  
\end{equation} Now, the two Shockley partials lying on the twin boundary ($C \delta$ and $\delta A$), which are separated by a finite distance, 
constitutes an extended dislocation with a Burgers vector of 1/2[01$\bar{1}]$ or CA, which also lies on twin boundary (Figure \ref{112-compression1}c).
With increasing deformation, the incident extended dislocation completely cross-slips to the twin boundary plane and glides further 
on the same plane as shown in Figure \ref{112-compression1}d.

Thus, under the compressive loading of twinned nanopillars, the extended dislocations show symmetric slip transmission (Figure 
\ref{112-compression}) and also the cross-slip to the twin boundary plane (Figure \ref{112-compression1}).

\subsection{Deformation mechanisms in twinned $<$112$>$ Cu nanopillars under tensile loading}

\begin{figure}[h]
\centering
\includegraphics[width=12cm]{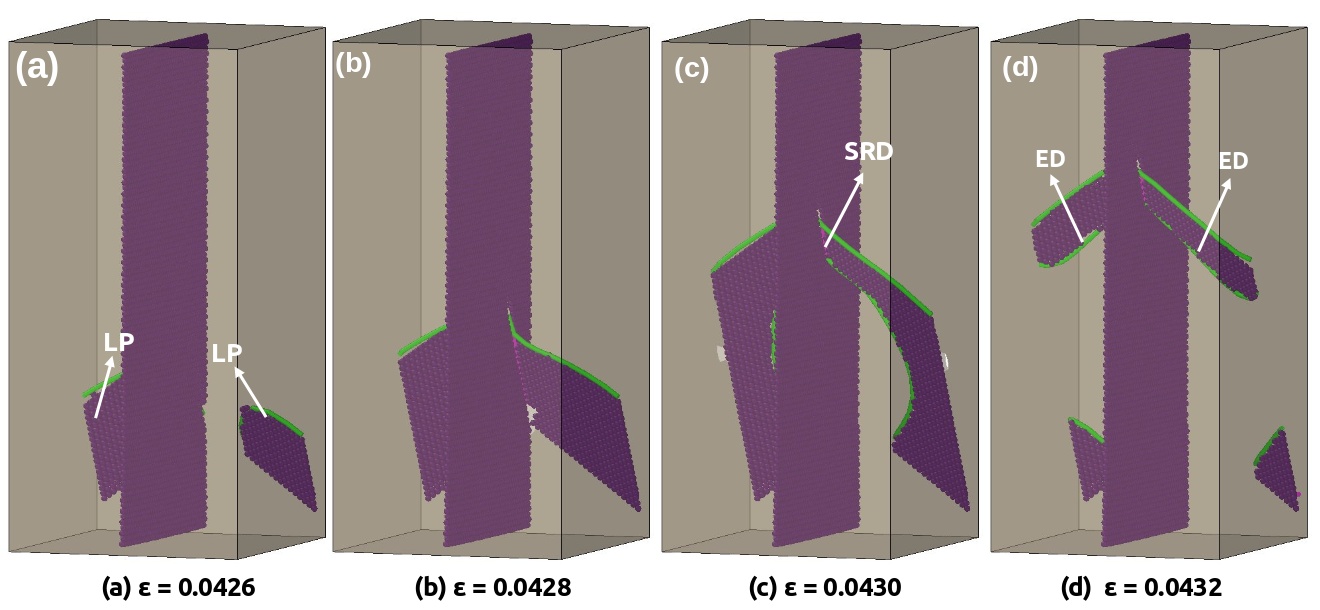}
\caption {The atomic configurations displaying the deformation behaviour of twinned Cu nanopillars under tensile loading. 
The two leading partials \hl{(LP)} (a-b) interacts at the twin boundary and results in the formation of a stair-rod dislocation (b).
The stair-rod dislocation \hl{(SRD)} again dissociates into two trailing partials (c), which glide on the same plane as that of the 
leading partials but in opposite direction. The sequential glide of leading and trailing partials enclosing the stacking 
fault makes the deformation to proceed by extended dislocations \hl{(ED)} (d). The purple colour surface represents the stacking fault 
and twin boundary. The green colour lines indicate the Shockley partial dislocations and the magenta colour line represents 
a stair-rod dislocation.}
\label{112-tension}
\end{figure}

Like compressive loading, tensile deformation of nanopillars with single and multiple twin boundaries has been investigated. 
Figure \ref{112-tension} shows the atomic snapshots of twinned $<$112$>$ Cu nanopillar containing a single twin boundary under 
tensile loading. The yielding in twinned nanopillar occurs through the nucleation of leading partial dislocations from two 
different corners of the nanopillar (Figure \ref{112-tension}a). The two nucleated leading partials glide towards each other and 
interact at twin boundary resulting in the formation of a stair-rod dislocation (Figure \ref{112-tension}b), which again lies 
on twin boundary plane. With increasing deformation, the stair-rod dislocation again dissociates into two trailing partials 
gliding on the same planes as that of the original leading partials (Figure \ref{112-tension}c). The dissociation of stair-rod 
into two trailing partials annihilate the stacking faults produced by leading partials (Figure \ref{112-tension}c). This novel 
mechanism of stair-rod formation and dissociation results in deformation proceeding through extended dislocations (leading 
followed by trailing) in twinned nanopillars as shown in Figure \ref{112-tension}d. Similar formation of stair-rod and 
its dissociation at the twin boundary has been reported in earlier studies \cite{SainathPLA2015,ZhangPLA2016}.

\begin{figure}[h]
\centering
\includegraphics[width=10cm]{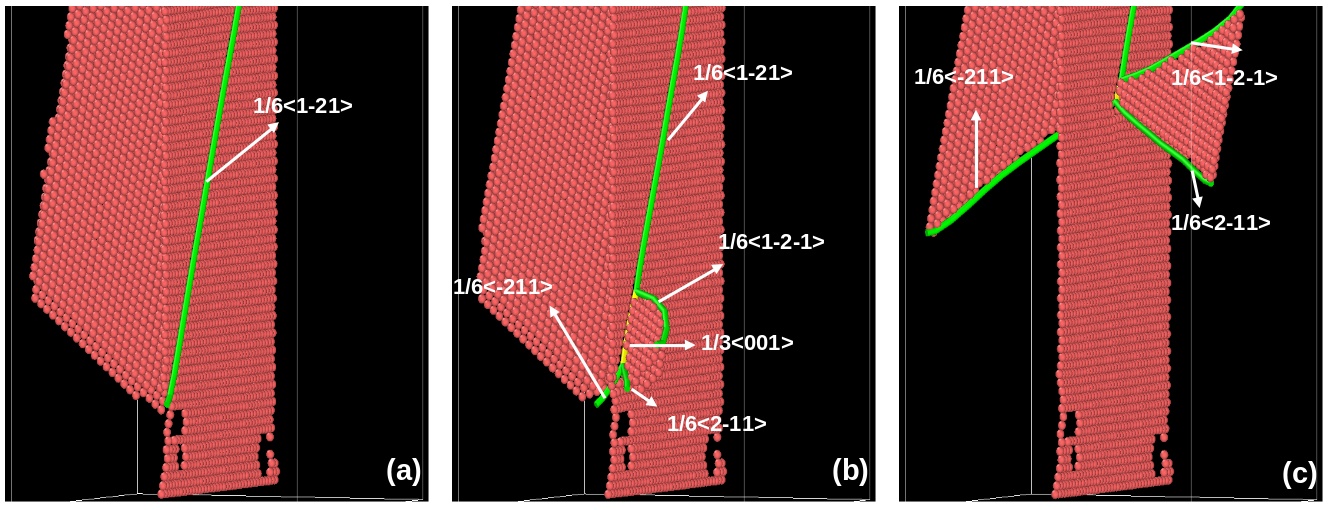}
\caption {The deformation mechanism under the tensile loading of twinned Cu nanopillars. The leading partial impeded by the twin 
boundary can be seen in (a). This partial dislocation cross-slips to the next grain leaving behind a Hirth stair-rod dislocation 
at the twin boundary (b). The Hirth dislocation again dissociates into two partials, which results the deformation in the nanopillar 
to proceed by extended dislocations (c). The perfect and surface atoms are removed for clarity. The red colour atoms represents the 
HCP atoms (stacking fault and twin boundaries). The green colour lines indicate Shockley partials and yellow colour line shows 
Hirth stair-rod.}
\label{112-tension1}
\end{figure}

In addition to stair-rod formation through two leading partials at the twin boundary, the stair-rod formation due to a single leading 
partial dislocation has also been observed as shown in Figure \ref{112-tension1}. Here, once the leading partial encounters the twin boundary 
(Figure \ref{112-tension1}a), it cross-slips to the next grain leaving behind a Hirth stair-rod dislocation at the twin boundary (Figure 
\ref{112-tension1}b) according to the following reaction, 
\begin{equation} 
\label{eq:00}
{1\over6} [1\bar{2}1] \longrightarrow {1\over6}[1\bar{2}\bar{1}] + {1\over3}[001] \quad \text{or} \quad \alpha C \longrightarrow \alpha ' B + \alpha \alpha'/CB
\end{equation}With increasing strain, the Hirth dislocation on the twin boundary dissociates into two Shockley partial dislocations, 
one gliding back into the original grain and the other gliding behind the leading partial on the cross-slip plane (Figure \ref{112-tension1}b).
This dissociation of Hirth dislocation can be written as 
\begin{equation} 
\label{eq:01}
{1\over3}[001] \longrightarrow {1\over6}[2\bar{1}1] + {1\over6}[\bar{2}11] \quad \text{or} \quad \alpha \alpha'/CB \longrightarrow B\alpha + C\alpha'
\end{equation}Here, it can be seen that these dissociated partials act as trailing dislocations making the deformation in nanopillars 
proceed through extended dislocations (Figure \ref{112-tension1}b-c). The Burgers vector of these extended dislocations is the summation 
of leading and trailing partials as shown below
\begin{equation} \label{eq:02}
{1\over6}[1\bar{2}\bar{1}] + {1\over6}[2\bar{1}1] \longrightarrow {1\over2}[1\bar{1}0] \quad \text{or} \quad C\alpha' + \alpha'B \longrightarrow CB
\end{equation} 

\begin{figure}[h]
\centering
\includegraphics[width=11cm]{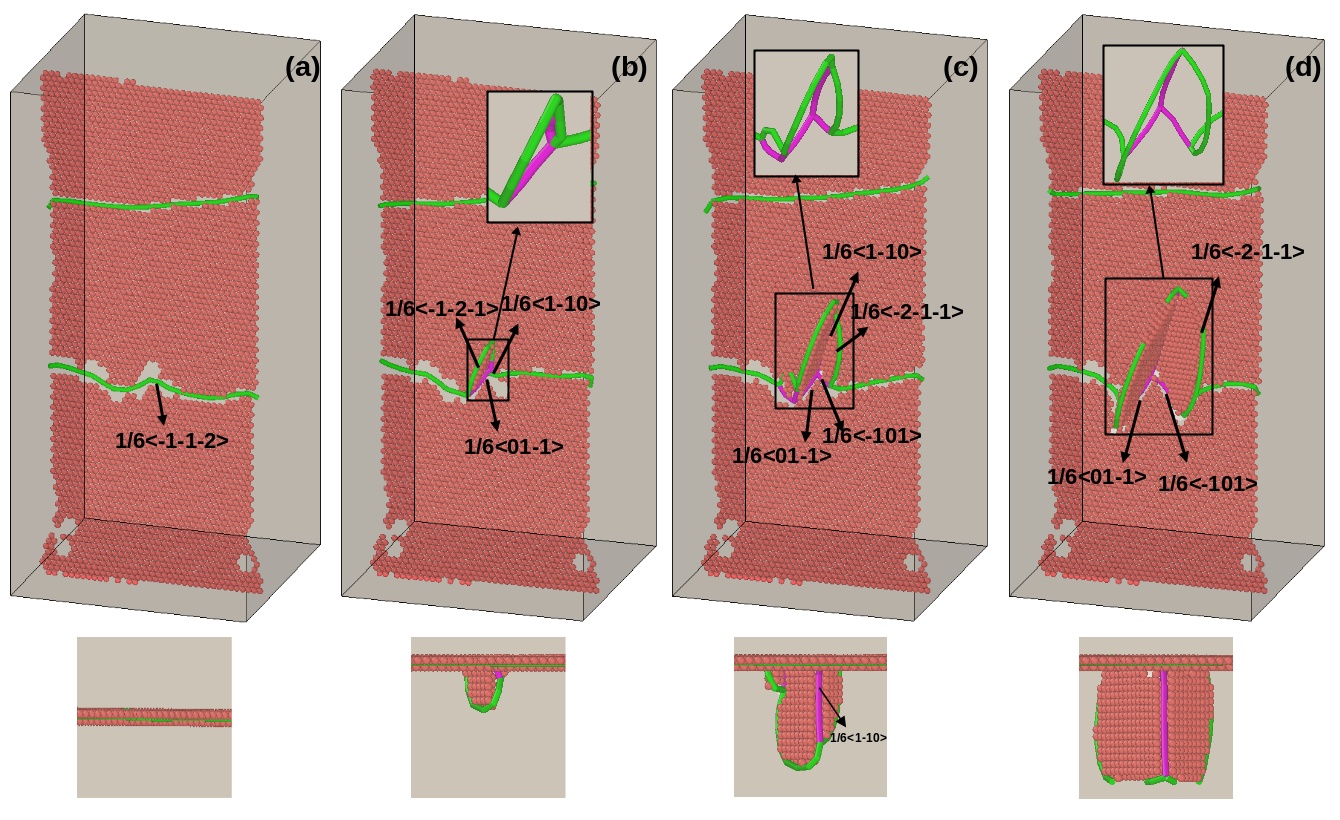}
\caption {The dislocation activity at the twin boundary resulting in the formation of well organized dislocation network. The bottom 
line figure shows the top view of the nanopillar. For clarity, the dislocation network without the stacking fault atoms are shown in 
inset figures.}
\label{triangularprism}
\end{figure}

Similar to the cross-slip mechanism seen in compressive loading, it has been observed that under tensile loading also the 
extended dislocation can cross-slip to twin boundary plane. However, this cross-slip onto the twin boundary plane has been mediated 
by forming a constriction. 
Following the cross-slip, the extended dislocation lying on the twin boundary can be seen in Figure \ref{triangularprism}a. Interestingly, 
one partial of this extended dislocation gets dissociated into a stair-rod dislocation (which lies on the twin boundary) and a Shockley 
partial dislocation (which protruded out of the twin boundary) as shown in Figure \ref{triangularprism}b. This dissociation can be written 
as 
\begin{equation} 
\label{eq:12}
{1\over6} [\bar{1}\bar{1}\bar{2}] \longrightarrow {1\over6}[01\bar{1}] + {1\over6}[\bar{1}\bar{2}\bar{1}] \quad \text{or} \quad \delta C \longrightarrow \delta \alpha '+ \alpha'C
\end{equation} Following this dissociation, the small segment of protruded Shockley partial dislocation again dissociates into a stair-rod 
dislocation and a Shockley partial dislocation as shown in Figure \ref{triangularprism}c. This dissociation in Burger's vector terms can be 
written as 
\begin{equation} 
\label{eq:13}
{1\over6} [\bar{1}\bar{2}\bar{1}] \longrightarrow {1\over6}[1\bar{1}0] + {1\over6}[\bar{2}\bar{1}\bar{1}] \quad \text{or} \quad \alpha' C \longrightarrow \alpha ' \beta' + \beta'C
\end{equation} As the deformation proceeds, the length of both stair-rod dislocation and Shockley partial dislocation increases 
(Figure \ref{triangularprism}c). Then, a new stair-rod dislocation (which lies on the twin boundary) forms upon the interaction of 
dissociated Shockley partial and a Shockley partial lying on the twin boundary (Figure \ref{triangularprism}c). The formation of 
stair-rod occurs according to the following reaction:  
\begin{equation} 
\label{eq:14}
{1\over6} [\bar{2}\bar{1}\bar{1}] + {1\over6}[112]  \longrightarrow  {1\over6}[\bar{1}01] \quad \text{or} \quad \beta'C + C\delta \longrightarrow \beta'\delta
\end{equation}Upon new stair-rod formation on the twin boundary, all dislocations (stair-rods and Shockley partials) lying on the twin 
boundary plane glide towards the surface (Figure \ref{triangularprism}d). Similarly, the protruded stair-rod and Shockley partial 
dislocations also move away from the twin boundary (Figure \ref{triangularprism}d). 

\begin{figure}[h]
\centering
\includegraphics[width=6cm]{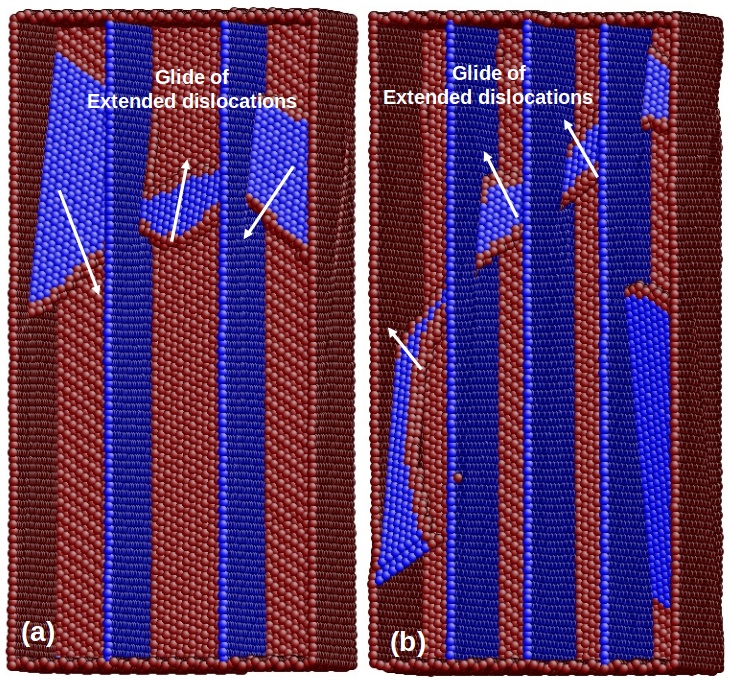}
\caption {The atomic snapshots displaying the deformation behaviour of twinned  $<$112$>$ Cu nanopillars under tensile loading. 
The deformation behaviour in nanopillar with (a) two twins and, (b) three twins. The atoms are coloured according to the
common neighbour analysis (CNA). The blue colour atoms represents the FCC and the red colour atoms indicate the surfaces 
and dislocation core.}
\label{112-tensionmultiple}
\end{figure}

Thus, the stair-rod formation due to two or one leading partials makes the deformation to proceed by extended dislocations under 
tensile loading. Likewise, Figure \ref{112-tensionmultiple} shows the plastic deformation in nanopillar with multiple twin boundaries, 
which is also dominated by extended dislocations formed through similar mechanism of stair-rod formation and dissociation. Here, it is 
interesting to note that the plastic deformation in twinned nanopillars (which occurred through extended dislocations) is quite different 
from partial dislocation slip in perfect $<$112$>$ nanopillars.

\section{Discussion}

\subsection{Yield stress in twinned nanopillars}

The results with respect to the strength of axially twinned Cu nanopillars indicate that, irrespective of twin boundary spacing, all the 
twinned nanopillars possess higher strength as compared to their perfect counterparts (Figure \ref{Fig03}). Further, the yield strength 
decreases with increasing the twin boundary spacing. In orthogonally twinned nanopillars, the higher strength of twinned nanopillars has 
been attributed to different factors like redistribution of interior stress within the nanopillar \cite{Cao2007} and strong repulsive 
force offered by twin boundaries on dislocations \cite{Chen2007,Deng-Scripta,Guo-Acta}. In axially twinned nanopillars also similar 
factors influence the yield strength. 

\begin{figure}[h]
\centering
\begin{subfigure}[b]{0.48\textwidth}
\includegraphics[width=\textwidth]{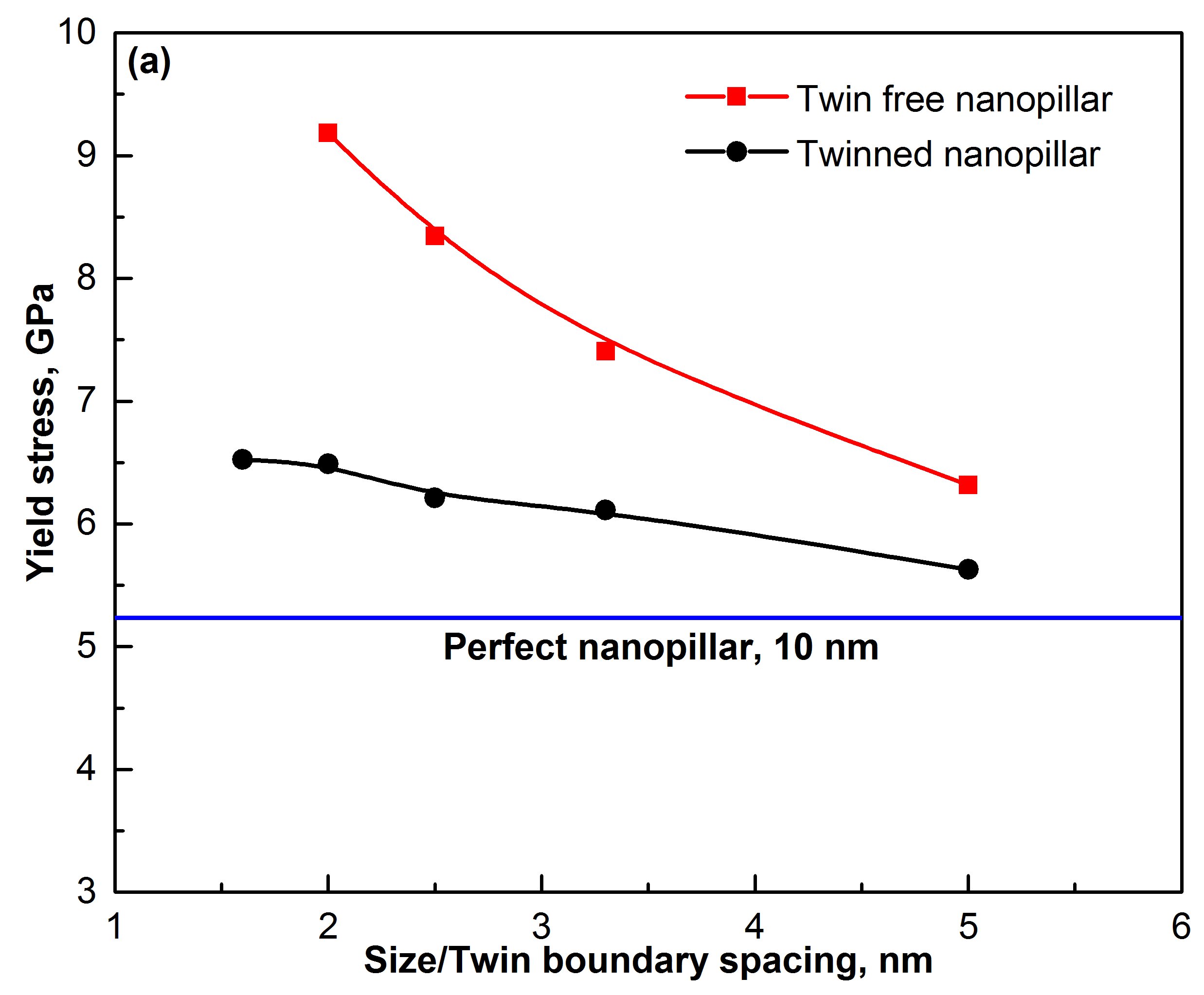}
\end{subfigure}
\begin{subfigure}[b]{0.48\textwidth}
\includegraphics[width=\textwidth]{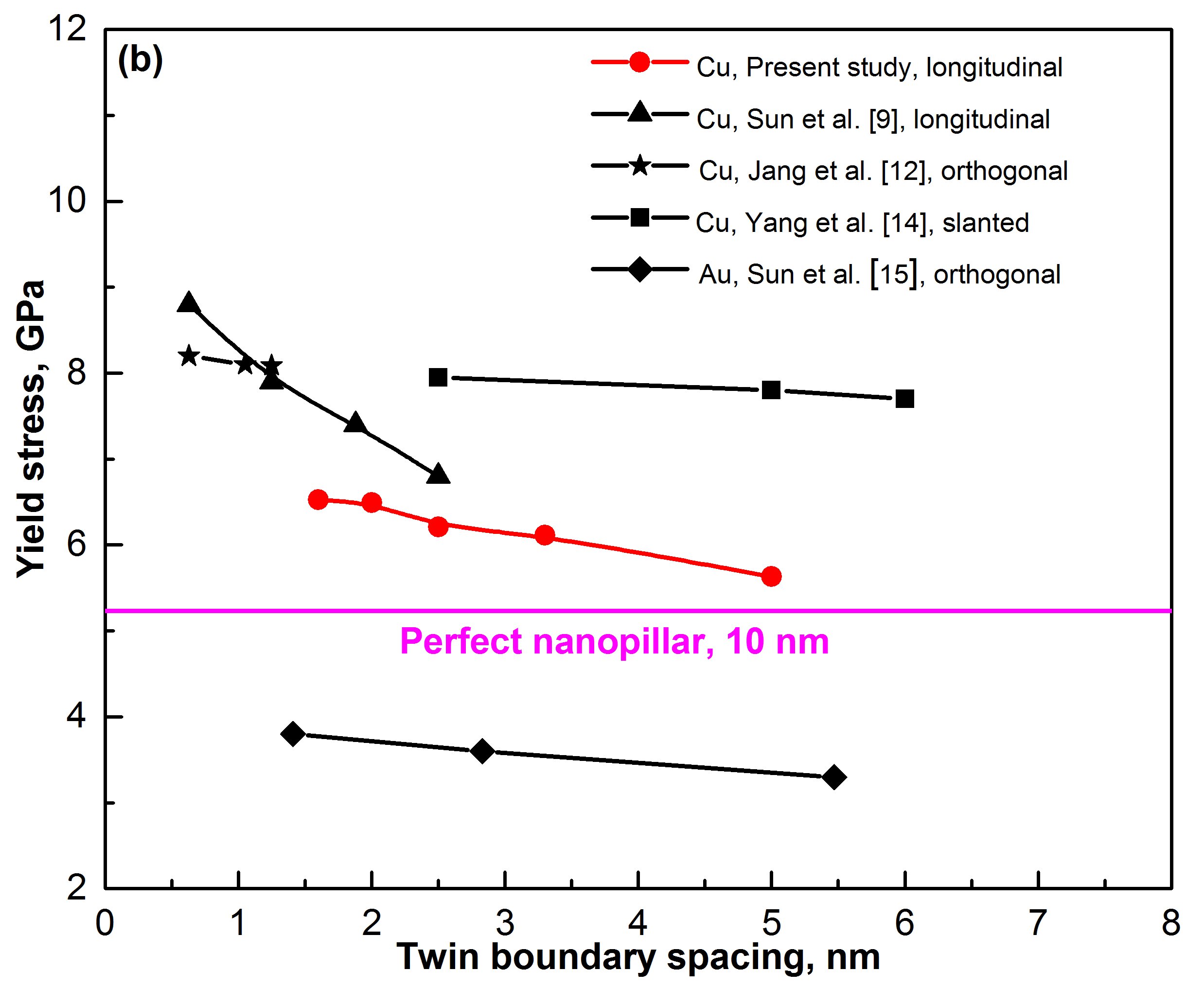}
\end{subfigure}
\caption {\hl{(a) Comparison of yield stress in twinned and twin free nanopillars with size equal to twin boundary spacing. 
(b) Comparison of yield stress in twinned nanopillars with that reported in other studies from the literature \mbox
{\cite{Sun2017,Jang2012,Yang2017,Sun2018}}. For comparison, the yield stress of perfect nanopillar of size 10 nm has also 
been shown.}}
 \label{Size-TBS}
 \end{figure}

In order to understand 
the role of stress redistribution in individual twinned regions, the strength of single crystal nanopillars with cross-section width (d) 
equal to the twin boundary spacing has been investigated. Figure \ref{Size-TBS}a shows the variation of yield strength in perfect nanopillars 
as a function of cross-section width (d) (here d is equivalent to twin boundary spacing). For comparison, the strength of original perfect 
nanopillar (d = 10 nm) and twinned nanopillars of different twin boundary spacing has also been shown in Figure \ref{Size-TBS}a. It can be seen
that, the perfect nanopillars (of size equivalent to twin boundary spacing) exhibit an \hl{expected size effect} i.e., strength decreases with 
increasing size \cite{Rohith-Philmag}. Further, the yield strength of the twinned nanopillar is much lower than the yield strength of 
corresponding perfect nanopillar with 
size equal to twin boundary spacing. With this comparison, it can be concluded that the strength of individual twinned regions has some 
contribution in dictating the overall strength of the twinned nanopillars. However, it is not directly additive i.e., the sum of the yield 
strength of individual perfect nanopillars (of size equal to twin boundary spacing), is not equal to the strength of twinned nanopillar of 
same spacing. This may be attributed to the surface modification introduced due to the incorporation of twin boundaries from individual 
perfect nanopillars.

The other factor is the repulsive force acting on the dislocations due to twin boundaries \cite{Sainath-PhilMag,Chen2007,Deng-Scripta,
Guo-Acta}. Deng and Sansoz \cite{Deng-Scripta} have shown that the repulsive force acting on the dislocations is proportional to the 
Burgers vector and inversely proportional to twin boundary spacing. \hl{Generally, the critical resolved shear stress (CRSS) for dislocation 
nucleation in twinned nanopillars ($\tau_c$) can be written as \mbox{\cite{CMS-19}} $\tau_c = \tau_o + \tau_{tb}$, where $\tau_c$ is the CRSS in 
pure twin free nanopillar and $\tau_{tb}$ is the additional contribution to CRSS due to the presence of twin boundaries. Here, $\tau_o$ is 
simply the product of yield stress and maximum value of Schmid factor in perfect nanopillars and the $\tau_{tb}$ is given by Deng and 
Sansoz \mbox{\cite{Deng-Scripta}} as follows;}

\begin{equation} 
\tau_{tb} = \lambda {{\mu bsin\theta} \over {4\pi x (1- \nu)}}\left( 1+ {b \over 2r}ln{8r \over b} \right)
\end{equation} \hl{Here for Cu, $\lambda$ is approximately 0.3 \mbox{\cite{Deng-Scripta}}, the shear modulus ($\mu$) is 53.31 GPa, 
$\theta$ is the angle between the twin boundary and the \{111\} slip plane, $\nu$ is the Poisson’s ratio, $b$ is the Burgers vector of 
partial dislocations, $x$ is the twin boundary spacing, $r = \sqrt{bDcos \theta}$ and $D$ is the size of the nanopillars. Based on this 
formulation and given 
parameters, the values of $\tau_c$ have been calculated for nanopillars with different twin boundary spacing in the range 1.6 - 5 nm. It 
has been observed that $\tau_c$ varies in the range 2.28 GPa (for nanopillar with $x$ = 1.6 nm) to 2.11 GPa (for nanopillar with $x$ = 5 
nm).} In other words, higher is the twin boundary spacing lower is the repulsive force and this leads to lower yield stress values. In 
nanopillars with smaller twin boundary spacing, the high repulsive force makes the dislocation nucleation and glide difficult, thus giving 
rise to increase in yield strength.

\hl{Figure \mbox{\ref{Size-TBS}}b compares the yield stress values of twinned nanopillars in the present study with that reported in 
other studies in literature \mbox{\cite{Sun2017,Jang2012,Yang2017,Sun2018}}. Similar to the present study, many studies in the literature 
have reported that the yield stress decreases with increasing twin boundary spacing. However, the absolute values of yield stress are 
different in different studies. This difference in yield stress values may arise due to many factors like nanowire size, cross section 
shape and twin boundary orientation.}

\subsection{Deformation mechanisms and cross-slip}

The results also indicate that the axial twin boundaries significantly influence the deformation mechanisms of Cu nanopillars. The deformation 
in perfect $<$112$>$ nanopillars occurs by partial dislocation slip/twinning under tensile loading. Generally, in $<$112$>$ oriented nanopillars, 
the deformation by partial dislocation slip is predicted according to Schmid factor analysis \cite{Rohith-ComCondMater}. However, the 
introduction of twin boundaries changed the deformation mode from partial dislocation slip/twinning to extended dislocation slip. The extended 
dislocations in twinned nanopillars are facilitated by the stair-rod formation and its dissociation on the twin boundary. On the other hand, 
the deformation is dominated by the slip of extended dislocations under the compressive loading of perfect as well as twinned nanopillars, 
which is in agreement with Schmid factor based predictions \cite{Rohith-ComCondMater}. This indicate that under compression, the deformation 
mechanisms are not influenced by the presence of twin boundaries.

\begin{figure}[h]
\centering
\includegraphics[width=12cm]{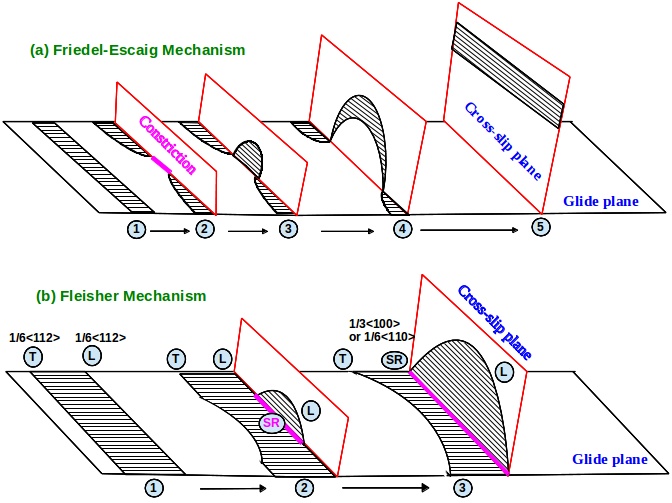}
\caption {The schematic view showing a sequence of events in (a) Friedel-Escaig (FE) \cite{Friedel1957,Friedel1964,Escaig1968}, and (b) 
Fleischer mechanism \cite{Fleischer1959} of cross-slip. In FE 
mechanism, the constriction of two partials in the initial glide plane followed by re-dissociation into the cross-slip plane can be seen. In 
Fleischer mechanism, the leading partial and stacking fault fold over from initial glide plane to cross-slip plane, by leaving a stair-rod 
dislocation at the intersection.}
\label{FE-FL}
\end{figure}

Irrespective of loading conditions, the presence of twin boundaries has introduced an extensive cross-slip of dislocations from one grain to 
the neighbouring grain and also the cross-slip onto the twin boundary plane. This indicates that the presence of twin boundaries significantly 
influences the deformation in FCC Cu nanopillars by the cross-slip activity. Since the deformation under compressive loading 
is dominated by extended dislocations, these extended dislocations get constricted at the twin boundary and results in formation of full 
dislocation (Figure \ref{112-compression}). This constricted full dislocation, which has high energy, cross-slip to neighbouring grain on a 
plane symmetric to the original by dissociating into an extended dislocation (Figure \ref{112-compression}). This mechanism is similar to 
the Friedel-Escaig (FE) \cite{Friedel1957,Friedel1964,Escaig1968} cross-slip mechanism as shown schematically in Figure \ref{FE-FL}a. The 
FE cross-slip mechanism involves the constriction of two partials in the initial glide plane followed by re-dissociation into the cross-slip 
plane (Figure \ref{FE-FL}a). In addition to FE mechanism, it has also been observed that the extended dislocation can also cross-slip to twin 
boundary plane without forming any constriction. As shown in Figure \ref{112-compression1}, the cross-slip of an extended dislocation is 
also facilitated by forming an intermediate stair-rod dislocation, which is similar to Fleischer mechanism of cross-slip in FCC metals 
\cite{Fleischer1959} as shown schematically in Figure \ref{FE-FL}b. In Fleischer mechanism of cross-slip, the leading partial dissociates 
into a stair-rod and another Shockley partial on the cross-slip plane without the concept of constriction. In other words, the leading 
partial and stacking fault fold over from initial glide plane to cross-slip plane, by leaving a stair-rod dislocation at the intersection 
(Figure \ref{FE-FL}b). If the the glide plane and cross-slip plane are at an acute angle, then the low energy stair-rod (1/6$<$110$>$) 
forms at the intersection (Figure \ref{112-compression1}), while high energy Hirth stair-rod (1/3$<$100$>$) forms at obtuse intersection 
(Figure \ref{112-tension1}).

Similar to compressive 
loading, cross-slip through Fleischer mechanism has also been observed under tensile loading (Figure \ref{112-tension1}). However, the 
Friedel-Escaig mechanism of cross-slip has not been observed. This is mainly because the trailing partials under tensile loading are 
generated from the 
dissociation of stair-rod dislocation lying on the twin boundary. This restricts the probability of constriction of leading and 
trailing partials. Similar to that observed in the present study, Zhu et al. \cite{Zhu2008} have modeled the slip transfer reactions of 
$1/2<110>$ screw dislocations in nanotwinned Cu bi-crystals and predicted the two different cross-slip mechanisms in accordance with 
Friedel-Escaig and Fleischer. Further, Fleischer mechanism was also observed  in atomistic simulations of stress driven cross-slip in 
FCC Al \cite{Vegge2000}.

The other interesting observation in the present study is the glide of partial/extended dislocation along the plane of twin boundary. Since 
the twin boundary is parallel to the loading direction, the shear stress acting on dislocations lying on this plane is zero. Therefore, the 
glide of any dislocations on the axial twin boundary is not expected. However, under both tensile and compressive loading, the frequent 
glide of partial/extended dislocations has been observed along this plane. An example of such glide can be seen from Figures 
\ref{112-compression1}c-d. Similar glide of partial dislocations migrating on the twin boundary has been reported by Jeon and Dehm 
\cite{Jeon-Dehm}. This unusual glide may be due to the rotation of the sample under tensile/compressive loading or stress localization in 
the vicinity of twin boundaries. 

\hl{Finally, it is important to mention that the twin boundary spacing considered in the present study is in the range 1.5 - 5 nm. This 
is in the same order of magnitude as that of the equilibrium spacing of the partial dislocations in Cu, which varies in the range 1.5 - 3 
nm, depending on the size \mbox{\cite{Rohith-AIP}} and 2 - 5 nm depending on the dislocation orientation \mbox{\cite{Cockayne}}. For 
a perfect nanowire of size 10 nm, the equilibrium spacing will be around 3 nm \mbox{\cite{Rohith-AIP}}. However, in the presence of twin 
boundaries, it may get reduced due to the repulsive force from twin boundaries. In this context, it is interesting to understand the 
relevance of this length scale to the observed deformation mechanisms, particularly the cross-slip. It is well known that the cross-slip 
mechanism strongly depends on the stacking fault energy (SFE) of the materials, which in turn is related to the equilibrium separation of 
partials. The dislocations easily cross-slips in materials with high SFE, while the cross-slip is difficult in low SFE materials. As the 
SFE is inversely proportional to equilibrium separation of partials, the extended dislocations having a wider partial separation finds it 
difficult to cross-slip, while cross-slip is easy for dislocations with smaller partial separation. Thus, it can be concluded that the 
length scale of the equilibrium separation of partial dislocations has some effect on the observed deformation mechanisms in the present 
study, particularly the cross-slip mechanism.}

\section{Conclusions}

The role of twin boundaries on the strength and deformation mechanisms of perfect and twinned Cu nanopillars has been investigated 
under tensile and compressive loading by means of atomistic simulations. The simulation results indicated that the presence of twin 
boundaries strengthens the nanopillars. The yield strength in longitudinally or axially twinned nanopillars increases with decreasing 
twin boundary spacing, which is similar to that observed in orthogonally twinned nanopillars. This strengthening in twinned nanopillars 
may be attributed to repulsive force offered by twin boundaries on dislocations and stress redistribution in individual twinned regions.
Further, the yield strength of twinned nanopillars show tension-compression asymmetry, which is higher than that exhibited by perfect 
nanopillars. 

In addition to strength, the presence of twin boundaries also changes the operative deformation mechanisms as compared 
to perfect nanopillars. Under compressive loading, the deformation in $<$112$>$ perfect as well as twinned nanopillars is dominated by 
the slip of extended dislocations. Further, the presence of twin boundaries under compressive loading introduces extensive cross-slip 
activity from one grain the other and also onto the twin boundary plane. This cross-slip activity under compressive loading occurs 
through two different mechanisms namely Friedel-Escaig mechanism and Fleischer mechanism. However, no cross-slip activity has been 
noticed in perfect nanopillars. 

On the other hand, under tensile loading, the deformation in perfect nanopillars occurs by partial 
dislocation slip/twinning, which changes to extended dislocation activity following the introduction of twin boundaries. This change 
in twinned nanopillars is due to the formation and dissociation of a stair-rod dislocation on the twin boundary. This stair-rod 
dislocation, which forms through the interaction of two leading partials at the twin boundary, dissociates into two trailing partials 
gliding on the same plane as that of the leading partials, thus constituting an extended dislocation slip. This extended dislocations 
under tensile loading also exhibit cross-slip activity in twinned nanopillars. However, this cross-slip occurs only through Fleischer 
mechanism and no Friedel-Escaig mechanism of cross-slip has been observed under tensile loading. Further, the dislocation glide along 
the twin boundary has also been observed under tensile and compressive loading, despite zero resolved shear stress on the twin boundary 
plane.

\end{document}